\documentclass{midl} 

\usepackage{mathrsfs}
\newcommand{\vect}[1]{\boldsymbol{#1}}

\jmlrproceedings{MIDL 2020}{Medical Imaging with Deep Learning 2020}
\jmlrvolume{}
\jmlryear{}
\jmlrworkshop{MIDL 2020 -- Short Paper}
\editors{}

\title[Motion prediction in free-breathing scans via a recurrent multi-scale encoder decoder]{Spatiotemporal motion prediction in free-breathing liver scans via a recurrent multi-scale encoder decoder}

\midlauthor{\Name{Liset V{\'a}zquez Romaguera\nametag{$^{1}$}} \Email{liset.vazquez@polymtl.ca}\\
\Name{Rosalie Plantefeve\nametag{$^{2}$}} \Email{rosalie@plantefeve.fr}\\
\Name{Samuel Kadoury\nametag{$^{1,2}$}} \Email{samuel.kadoury@polymtl.ca}\\
\addr $^{1}$ MedICAL Laboratory, Polytechnique Montreal, Montreal, Canada \\
\addr $^{2}$ CHUM Research Center, Montreal, Canada}

\begin{document}
\maketitle
\begin{abstract}
In this work we propose a multi-scale recurrent encoder-decoder architecture to predict the breathing induced organ deformation in future frames. The model was trained end-to-end from input images to predict a sequence of motion labels. Targets were created by quantizing the displacement fields obtained from deformable image registration. We report results using MRI free-breathing acquisitions from 12 volunteers. Experiments were aimed at investigating the proposed multi-scale design and the effect of increasing the number of predicted frames on the overall accuracy of the model. The proposed model was able to predict vessel positions in the next temporal image with a mean accuracy of $2.07 \pm 2.95$ mm showing increased performance in comparison with state-of-the-art approaches.
\end{abstract}

\section{Introduction}
According to the last American Cancer Society's report \citep{acs2020}, about 42,810 new cases of primary liver cancer will be diagnosed this year in the US. Radiation therapy is the first line of treatment for the majority of these cases. Its goal is to focus the radiation beams in the target and to avoid surrounding anatomy. However, respiratory motion is one of the major issues with large dosimetric impact \cite{mechalakos2004}. Image-guided radiation treatments can greatly benefit from future frame prediction models since the beam can be re-positioned compensating for motion. State-of-the-art approaches rely on statistical \cite{samei2012,preiswerk2014} and biomechanical modeling \citep{nguyen2008adapting, fuerst2014patient}. Recently, deep learning-based solution have been reported for motion prediction in the medical context \cite{wang2018feasibility, teo2018feasibility, krebs2019probabilistic}. In this work, we propose a recurrent multi-scale encoder-decoder framework to perform in-plane spatio-temporal motion prediction from sequential images. 

\section{Method}
\begin{figure}[t]
    \centering
    \includegraphics[width=1.0\textwidth]{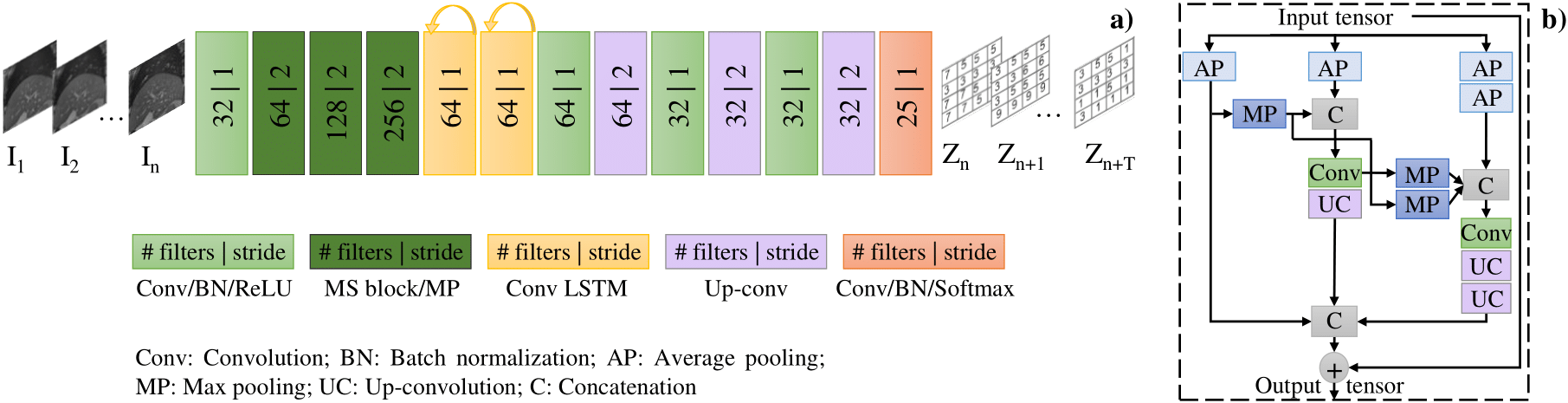}
    \caption{a) Proposed motion prediction architecture (b) Proposed multi-scale block.}
    \label{fig:model}
\end{figure}

The proposed model aims at learning a representation that predicts the sequence of encoded motion $\left\langle\vect{Z_n}, \vect{Z_{n+1}}, \ldots, \vect{Z_{n+T}}\right\rangle$ over $T$ future time steps given an input image sequence $\left\langle\vect{I_1}, \vect{I_2}, \ldots, \vect{I_n}\right\rangle$ of length
$n$. Consecutive pair of images were non-rigidly registered applying the B-spline transformation model and optimization based on normalized mutual information as implemented by NiftyReg software \cite{modat2010fast}. The resulting two-dimensional displacement fields were encoded using an auxiliary representation space $\vect{Z_i}$ = $\mathscr{F}(\vect{Y_i})$ where $\vect{Z_i} \in \mathbb{R}^{H \times W \times Q}$ and $Q$ is the number of classes. $\mathscr{F}$ is a mapping function to encode the displacement fields into labels. To that end, the ranges of values for each vectorial component, i.e. axes $x$ and $y$, were quantized into $b=5$ bins. A codebook $\textbf{C} \in \mathbb{R}^{Q} (Q=b^2)$ was built by assigning a class to each possible combination between the bins of each axis. As the probability distribution of the motion vectors obtained from deformable registration has an approximately-Gaussian shape, bins were selected near to the mean, standard deviation, minimum and maximum distribution values. Using this scheme, we effectively represent the motion observed in the dataset. The motion learning architecture is composed by a multi-scale (MS) encoder, recurrent units and a decoder, as illustrated in Figure~\ref{fig:model} (a). The MS encoder extracts feature representations at multiple scales through the network: fully, medium and low resolution in order to fully exploit the image features (see Figure~\ref{fig:model} (b)). The motion learning architecture also contains recurrent units and a fully convolutional spatial decoder. The spatio-temporal features extracted by the MS encoder are extrapolated in time by the convolutional Long Short-Term Memory (LSTM) units and further processed by the spatial decoder to recover the desired dimensions in the form of motion labels. We used a weighted cross entropy loss function as proposed by \cite{zhang2016colorful} to promote class rebalancing since the distribution is strongly biased toward classes representing the superior-inferior motion. Adam optimizer with an initial learning rate of $10^{-3}$ was used. This learning rate was reduced by 2 after 10 epochs without improvements in the validation set accuracy.

\section{Results and discussion}
\begin{figure}[ht]
   \begin{minipage}{0.49\textwidth}
     \centering
     \includegraphics[width=.95\linewidth]{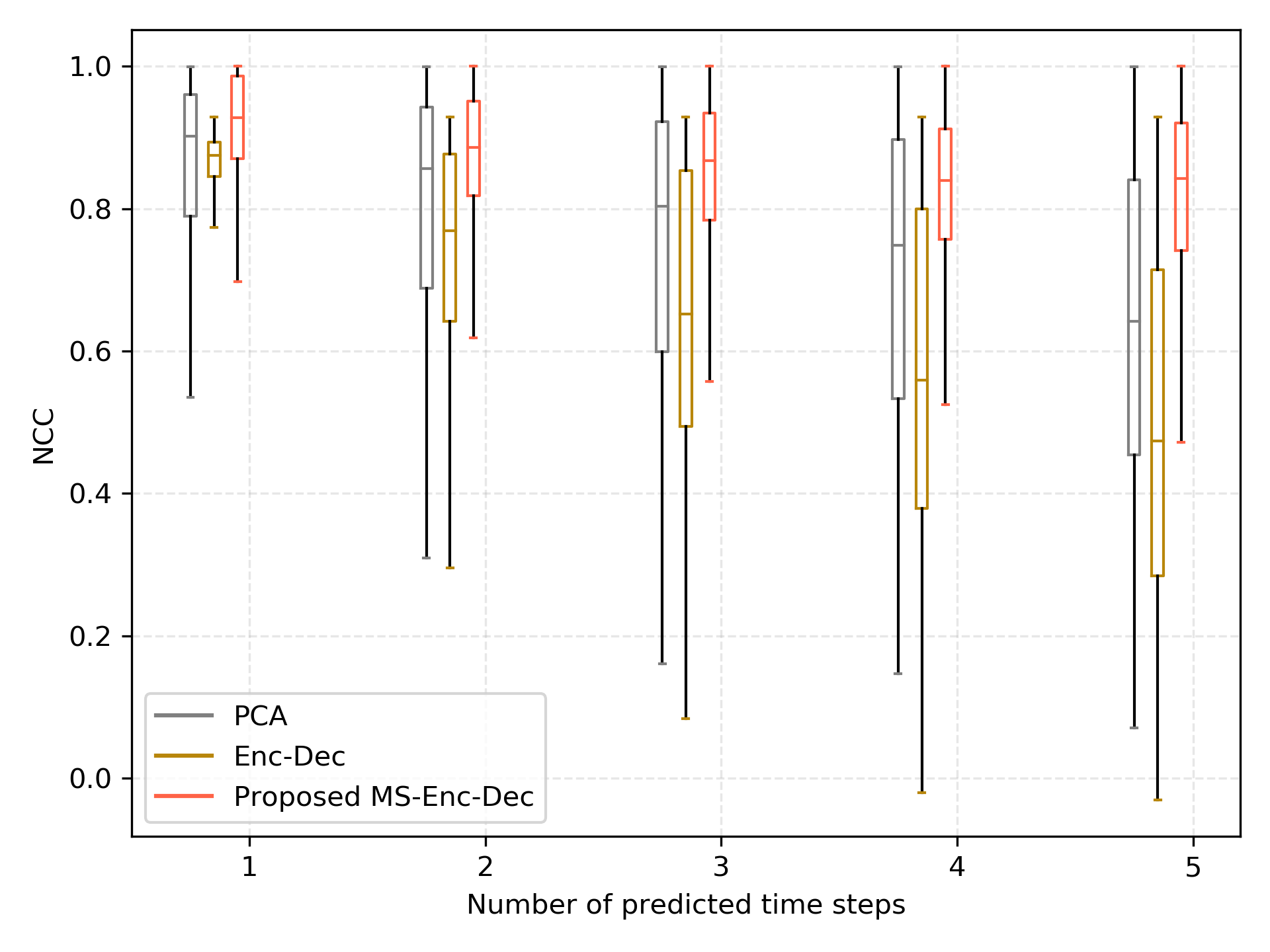}
     \caption{NCC between target and predicted images.}\label{fig:ncc}
   \end{minipage}\hfill
   \begin{minipage}{0.49\textwidth}
     \centering
     \includegraphics[width=.95\linewidth]{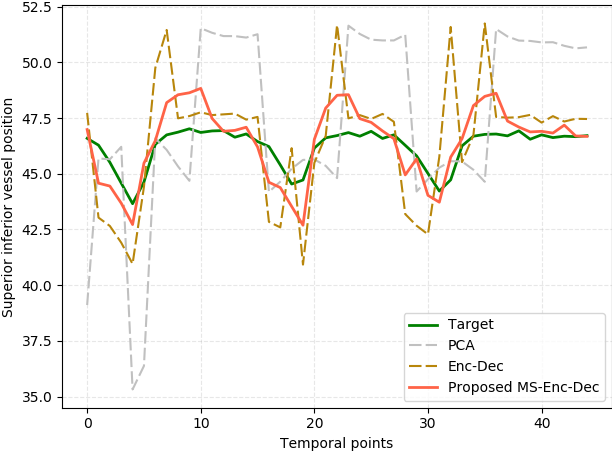}
     \caption{Vessel trajectory predicted with different approaches.}\label{fig:trajectories}
   \end{minipage}
\end{figure}

\begin{figure}[ht]
    \centering
    \includegraphics[width=0.9\textwidth]{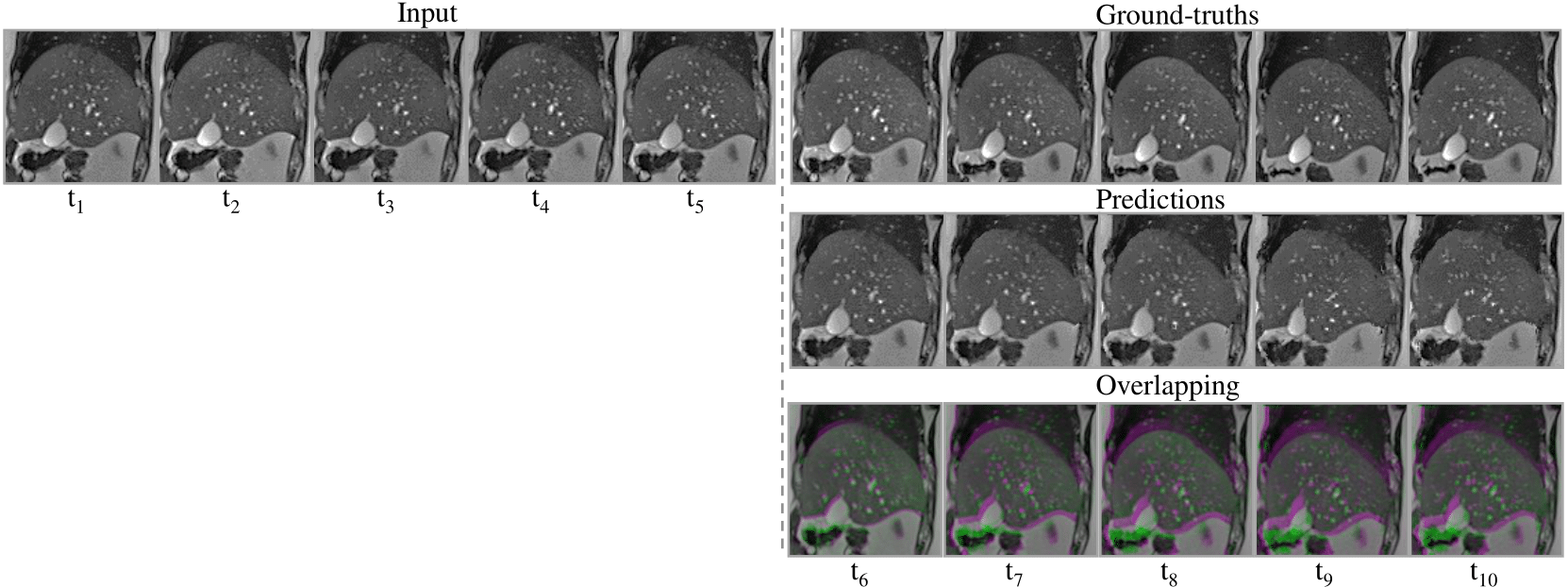}
    \caption{Extrapolation up to five time steps from an input sequence in the test set.}
    \label{fig:maps}
\end{figure}

Experiments were conducted using sagittal slices covering 15 positions on the right liver lobe from 12 volunteers. Each anatomical position comprises 50 dynamics. These cine-phase scans were acquired on a Siemens Skyra 3T scanner using a 2D T2-weighted true FISP sequence. Pixel spacing, slice thickness and temporal resolution are equal to $1.7 \times 1.7$ mm$^2$, 3 mm and 320 ms, respectively. The dataset was divided in train, validation and test sets following a leave-one-out scheme on a subject level. Thus, the model was tested on all the slice positions belonging to an unseen subject. We compared the proposed network with statistical modeling \cite{li2011pca} and with a similar architecture which uses the traditional feature extraction scheme (Conv-Pool stacking) \citep{luo2017unsupervised}. Two blood vessels were manually annotated in each image. We report results when varying the number of extrapolated times $\left\lbrace1, 2, 3, 4, 5\right\rbrace$ given 5 input images. Table~\ref{tab:vessel_tracking} presents mean Euclidean distances between ground truth and predicted vessel positions. Figure~\ref{fig:ncc} shows a comparison based on the Normalized Cross Correlation (NCC) metric. The proposed model outperforms the compared methods for the in-plane motion prediction task. Results show a lower performance when more time steps are predicted. It is natural that, based on the same information, the error increases when extrapolating more time points. Moreover, the model does not cope with out-of-plane motion which also influences the reported values. Figure~\ref{fig:trajectories} illustrates the vessel trajectory through the target and predicted temporal images. Our multi-scale encoder-decoder model showed the closest alignment with the target trajectory. Finally, Figure~\ref{fig:maps} shows an example of the output sequence obtained by deforming the last input image with the predicted deformations. While the model showed a competitive performance, some limitations should be considered. The first is related to the inherent error introduced by the quantization, which depends directly on the selected number of bins. Also, since the predicted displacement fields are recovered from motion labels, potential misclassification may lead to unrealistic and ambiguous motion. This aspect should be investigated in a future study.

\begin{table}[h]
\floatconts
  {tab:vessel_tracking}%
  {\caption{Vessel tracking error position (in mm) for each predicted time (mean $\pm$ std).}}%
  {\begin{tabular}{cccccc} 
  \bfseries Model & \bfseries t=1& \bfseries t=2& \bfseries t=3& \bfseries t=4& \bfseries t=5\\
  \bfseries     & \bfseries (320 ms)& \bfseries (640 ms)& \bfseries (960 ms)& \bfseries (1280 ms)& \bfseries (1600 ms)\\
  PCA & 2.25 $\pm$ 3.46 & 2.49 $\pm$ 3.76 & 3.45 $\pm$ 4.20 & 3.96 $\pm$ 4.42 & 4.39 $\pm$ 4.01\\
 Enc-Dec & 2.71 $\pm$ 3.21 & 3.41 $\pm$ 3.40 & 3.98 $\pm$ 4.17 & 4.41 $\pm$ 3.65 & 4.87 $\pm$ 4.11\\
  Proposed & 2.07 $\pm$ 2.95 & 2.24 $\pm$ 3.16 & 2.91 $\pm$ 3.52 & 3.11 $\pm$ 3.42 & 3.81 $\pm$ 3.63\\
  \end{tabular}}
\end{table}

\midlacknowledgments{This research was undertaken thanks, in part, to funding from the Canada First Research Excellence Fund through the TransMedTech Institute.}

\bibliography{mybib.bib}
\end{document}